# Spin-dependent ballistic transport and electronic structures in pristine and edge-doped zigzag silicene nanoribbons: large magnetoresistance

An-Bang Chen[1], Xue-Feng Wang[1*], P. Vasilopoulos[2], Ming-Xing Zhai[1], Yu-Shen Liu[3]

[1] Department of Physics, Soochow University, 1 Shizi Street, Suzhou 215006, China

[2] Concordia University, Department of Physics, 7141 Sherbrooke Ouest, Montreal, QC, Canada, H4B 1R6

[3] Jiangsu Laboratory of Advanced Functional materials, Changshu 215500, China

∗ xf_wang1969@yahoo.com

The electronic structure and conductance of substitutionally edge-doped zigzag silicene nanoribbons (ZSiNRs) are investigated using the nonequilibrium Green's function method combined with the density functional theory. Two-probe systems of ZSiNRs in both ferromagnetic and antiferromagnetic states are considered. Doping effects of elements from groups III and V, in a parallel or antiparallel magnetic configuration of the two electrodes, are discussed. Switching on and off the external magnetic field, we may convert the metallic ferromagnetic ZSiNRs into insulating antiferromagnetic ZSiNRs. In the ferromagnetic state, even- or odd-width ZSiNRs exhibit a drastically different magnetoresistance. In an *odd-width* edge-doped ZSiNR a large magnetoresistance occurs compared to that in a pristine ZSiNR. The situation is reversed in *even-width* ZSiNRs. These phenomena result from the drastic change of the conductance in the antiparallel configuration.

## 1. Introduction

Silicene, the graphene-like monolayer honeycomb structure of silicon, has been recently synthesized on Ag [1], ZrB2 [2], and Ir [3] substrate surfaces after the realization of silicene nanoribbons [4]. Though the origin of Dirac cone in the electronic band structure of silicene on Ag surface is still in question [5], it has been predicted that, similar to those of graphene, free-standing silicene has Dirac cones or a linear electronic energy dispersion near the Fermi energy [6,7]. Resulting from the large ionic radius of silicon atom, silicene has a buckled structure instead of the planar graphene structure. There exists a height difference between the two Si atoms in the primitive cell of silicene caused by the partial $sp^3$ characteristics rather than the complete $sp^2$ hybridization in graphene. In addition, silicene have a stronger spin-orbit interaction than that in graphene which might result in an energy gap around 1 meV [8]. The silicene's compatibility with silicon-based electronic technology suggests its advantage in potential device applications, see Ref. 9 for a recent review.

Graphene and graphene nanoribbons have been extensively investigated due to their properties for potential applications in nanodevices [10]. For the same reasons silicene [9] and silicene nanoribbons (SiNRs) [11] are attracting more and more interest. Electronic properties such as the spin-Hall effect [8], the anomalous Hall effect [12], the capacitance of an electrically tunable silicene device [13], the manipulation of band gap [14], and the doping effects [15,16] in silicene have been studied. The electron and phonon properties in SiNRs have been simulated by first principles calculation [17] and the giant magnetoresistance has been predicted in pristine even-width zigzag SiNRs (ZSiNRs) [18]. The effects of single and multiple dopants on the electric and magnetic properties of ZSiNRs have also been discussed [19].

In this work we simulate the spin-dependent ballistic transport properties of edge-doped $n$-ZSiNRs, with width $n$, [11,17-19] and compare them with their pristine counterparts [18]. The edge doping is produced by substituting a silicon (Si) atom on an edge of the ZSiNRs with a boron (B), nitrogen (N), aluminum (AL), or phosphorous (P) atom. We find that a large magnetoresistance appears only in odd-width edge-doped ZSiNRs, in contrast to the undoped ZSiNRs cases where it exists in even-width nanoribbons. We also highlight some differences with graphene nanoribbons. In Sec. 2 we present the model and in Sec. 3 the results obtained using the nonequilibrium Green's function (NEGF) method combined with density functional theory (DFT). Concluding remarks follow in Sec. 4.

## 2. Model and method

The geometric structure of a pristine *n*-ZSiNR with width *n*=4 is showed in Fig. 1. In our two-probe model used for the transport calculations, two ZSiNR electrodes (L and R) are attached to the ZSiNR device in the central region (C). In the doped cases we assume that the substitutionally doping atom locates at the center of the upper edge in region C. The length of the central region is chosen as 7 primitive cells, long enough to screen out the electrostatic effect of the doping atom on the electrodes. A vacuum layer, next to the two edges, thicker than 15 Å is used to eliminate possible mirror interaction and the edge silicon atoms are passivated by hydrogen atoms so as to eliminate the dangling bonds.

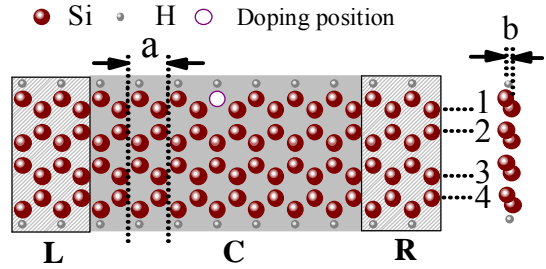

Fig. 1. (Color online) Top and side view of a 4-ZSiNR two-probe system, with left (L) and right (R) electrodes and 7 primitive cells in the center scattering region C. The red (gray) spheres are the Si (H) atoms. The unfilled sphere shows the Si atom that is replaced by a B, N, Al, or P atom in doped cases. The lattice constant is a = 3.87Å and the average vertical distance between the two Si sublattices b =0.57 Å.

The geometric optimization is carried out with the Atomistix ToolKits (ATK) package based on the density functional theory (DFT) in the generalized gradient approximation with the Perdew- Burke-Ernzerhof exchange-correlation functional. All structures are fully relaxed until the forces are smaller than 0.02 eV/Å on each atom.

The study focuses on the spin-dependent ballistic transport properties of sub-stitutionally edge-doped ZSiNRs by elements of the IIIA and VA groups. For that we use DFT combined with the NEGF formalism, as implemented in the ATK package [20,21], using the exchange-correlation functional in the local-density approximation with the Perdew-Zunger parametrization, a double-$\zeta$-polarization basis set, and a $1 \times 1 \times 500$ Monkhorst-Pack k-point mesh. The grid mesh cutoff is set to 250 Ry and the temperature of the electrodes to 300 K. The spin-dependent conductance is evaluated by the Landauer formula [20,21]

$$G_\sigma(E) = \frac{e^2}{h}T_\sigma(E) = \frac{e^2}{h}Tr[\Gamma_L G^R \Gamma_R G^A]_\sigma, \quad (1)$$

with $T_\sigma$ the transmission for spin $\sigma$, $\Gamma_L$ ($\Gamma_R$) the broadening matrix due to the left (right) electrode, and $G^R$ ($G^A$) the retarded (advanced) Green's function. The total conductance is $G(E) = \sum_\sigma G_\sigma(E)$ at energy $E$.

In each electrode of pristine ZSiNRs, the two edges can be spin polarized in the same and opposite directions referred to as the ferromagnetic (FM) and antiferromagnetic (AFM) state, respectively. Usually the AFM state is the ground state and the FM state can be the ground state under an external magnetic field as in the case of graphene zigzag nanoribbons [18,22]. For either FM or AFM electrodes as shown in Fig. 2 and Fig. 5, respectively, the magnetizations of the two electrodes can be aligned in a parallel (P) or antiparallel (AP) configuration. The magnetoresistance (MR) in the linear-response regime of systems with FM electrodes is then calculated using the definition [23]

$$MR^{FM} = \frac{G_P^{FM} - G_{AP}^{FM}}{\text{Min}\{G_P^{FM}, G_{AP}^{FM}\}}, \quad (2)$$

where $G_P^{FM}$ and $G_{AP}^{FM}$ are the total conductances at the Fermi energy $E_F$ of the two-probe system in the P and AP configuration of the electrodes, respectively. By switching on and off the external magnetic field, we can drive the systems from the FM state to the AFM state and change its conductance. The corresponding magnetoresistance between FM and AFM states in the P configuration can be defined as [24]

$$MR^B = \frac{G_P^{FM} - G_P^{AFM}}{\text{Min}\{G_P^{FM}, G_P^{AFM}\}}, \tag{3}$$

where $G_P^{AFM}$ is the total linear conductance of the system in the P configuration of AFM electrodes. Note that the spin orbit interaction (SOI) is not taken into account in the calculation. The SOI will open an energy gap about 1 meV in the FM states of ZSiNRs and can affect the linear conductance in real systems at low temperature.

## 3. Results and discussion
### A. *FM configuration*

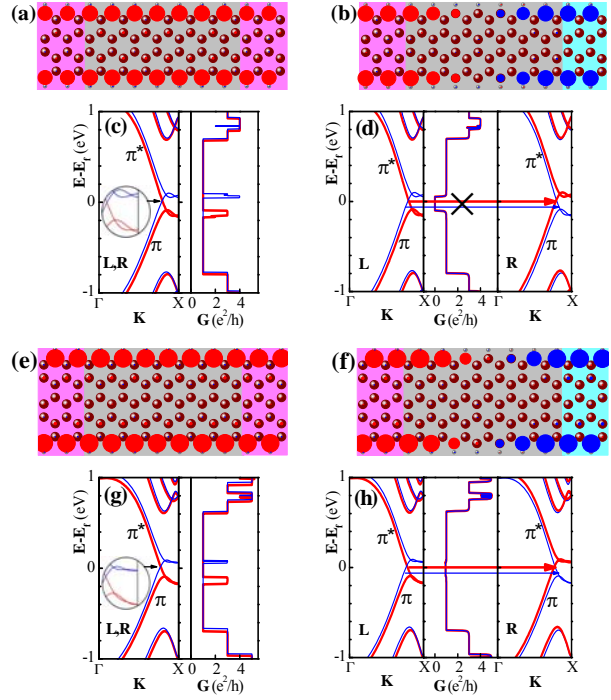

FIG. 2. Spin-polarization distribution in a two-probe system of pristine (a) FM-P 4ZSiNR, (b) FM-AP 4ZSiNR, (e) FM-P 5ZSiNR, and (f) FM-AP 5ZSiNR. The red (blue) filled circles on the atoms indicate the up (down) spin polarization and the magnitude of their radii the polarization difference between majority and minority electron spins on each atom. The corresponding energy bands for the left and right electrodes and the conductance under zero bias are shown in panels (c), (d), (g), and (h), respectively. The red thick curves are for spin-up and the blue thin ones for spin-down electrons. The energy bands near the Fermi energy are zoomed in the insets of (c) and (g). The cross sign in (d) indicates that there is no transport channel near the Fermi energy for both spins due to the orthogonality of the wave functions of the two electrodes.

First, we calculate the electron band structure and conductance of a perfect 4(5)-ZSiNR under zero bias in the FM state as presented in Fig. 2. Energies are measured from the Fermi level throughout the paper. The energy bands of opposite spins are close to each other except near the Fermi energy where the edge states are located. Every primitive cell of the 4(5)-ZSiNR possesses a net spin polarization as a whole. The magnetism comes mainly from the atoms on the edges as illustrated by the spin distribution presented in Fig. 2(a), (b), (e), and (f). This induces an effective magnetic field, in which electrons with spin opposite to it have higher energies, while those with spin parallel to it have lower energies as indicated by the energy bands in Fig 2.

In the P configuration the two-probe systems are periodic with translational symmetry and the conductance for each spin takes a step form; it is given by the number of transport channels times the conductance quantum $G_0 = e^2/h$. The energy bands of the edge states have twisted forms as shown by the insets in Fig. 2(c) and (g) for 4-ZSiNRs and 5-ZSiNRs, respectively. This may increase the number of transport channels at some energies and results in conductance peaks for spins up below (spins down above) the Fermi energy as shown in the right panels of Fig. 2(c) and (g). In the AP configuration, as shown in Fig. 2(d) and (h), the conductance spectra change only slightly at energies away from the Fermi energy due to the similarity of the energy bands of opposite spins. In contrast, the conductance changes drastically near the Fermi energy. Its peaks disappear because there is now at most one transport channel for each spin. In even-width ZSiNRs, due to their geometry symmetry, the wavefunctions of states in the π (π*) band, the band below (above) the twist as shown in Fig.2, are rotationally antisymmetric (symmetric) about the central line of the system [18]. At the Fermi energy, the spin-up (spin-down) band is the π* (π) band in the left electrode or the π (π*) band in the right electrode in the AP configuration. The electron wave functions of the π and π* bands are orthogonal to each other in even-width 4-ZSiNRs and the conductance drops greatly, from $G_0$ in the P configuration to almost zero in the AP one, as shown in Fig. 2(d). In odd-width 5-ZSiNRs the wave functions of the two electrodes are not orthogonal and the conductance is only slightly changed from $G_0$ when moving from one configuration to another. As a result, a giant $MR^{FM}$ occurs in 4-ZSiNRs but not in 5-ZSiNRs. [18]

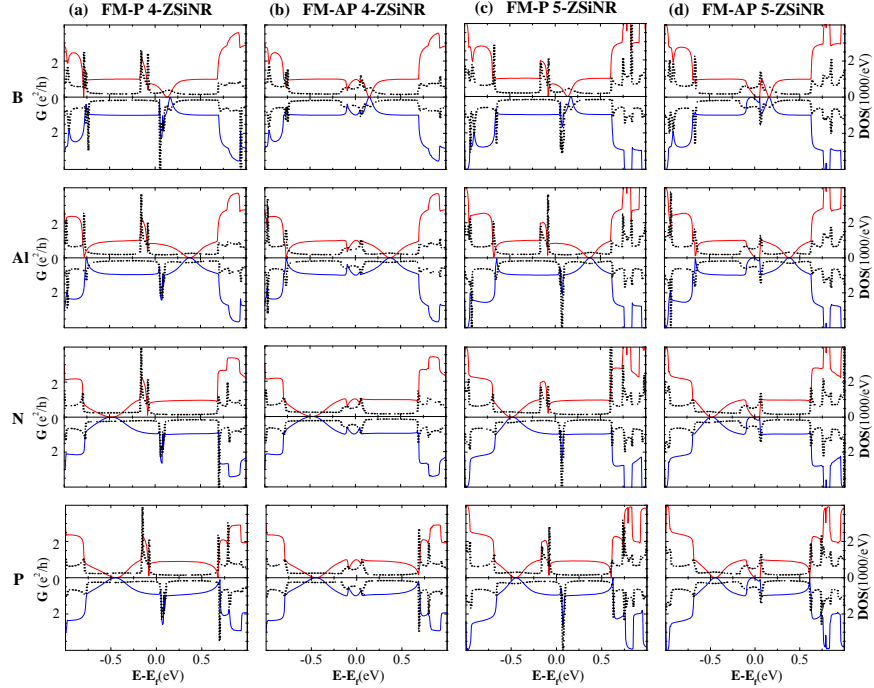

Fig. 3. Conductance and DOS of a FM 4- or 5-ZSiNR edge doped by a B, Al, N, or P impurity atom in P and AP electrode configurations. The solid red (blue) curve in the upper (lower) part of each panel is the spin-up (spin-down) conductance and the dotted black curve the corresponding DOS.

We now turn to the evaluation of the conductance $G$ of edge-doped 4(5)-ZSiNRs and the corresponding density of states (DOS). The Si atom, indicated by a red sphere in Fig. 1, is replaced by an impurity atom. In general, for electrons with energies near the Fermi energy, we have significant changes in the conductance of a pristine 4(5)-ZSiNR depending on the impurity atom and the P or AP electrode configuration. The details are as follows.

In Fig. 3 we show $G$ and the DOS for (a) a FM 4-ZSiNR in a P electrode configuration noted as FM-P 4-ZSiNR, (b) a FM-AP 4-ZSiNR, (c) a FM-P 5-ZSiNR, and (d) a FM-AP 5-ZSiNR edge doped by an atom of element B (first row), Al (second row), N (third row), and P (fourth row). The solid red (blue) curves present the spin-up (spin-down) conductances and the dotted black curves the spin-up (spin-down) DOSs. As can be seen, the results depend to some degree on the impurity atom, called X for convenience. Relative to Fig. 2 and noticing the way the spin-down results are plotted, we see new dips in the spin-up or spin-down $G$ occurring at energies of the localized impurity states that depend on X. In particular, for parallel and antiparallel electrodes, we highlight the spin-up or spin-down minima below $E_F$ near E = −0.5 eV for p-type doping elements, X = N, P, and the conductance minima above $E_F$ for n-type doping elements, X = B, near E = 0.1 eV and X = Al near E = 0.3 eV. For the heavier elements, minima of opposite doping type also appear like the ones near E = 0.6 eV, for X=P, and near E=−0.8eV for X=Al. Furthermore, the step form maxima of conductance near E=0.9 eV in Fig. 2 become round but are considerably less modified for every X. Interestingly, the conductance gap of FM-AP 4-ZSiNRs in the range E ∈ [-0.064, 0.064] eV in Fig. 2(d) disappears for every X. It is replaced in doped 4-ZSiNRs by a conductance peak of height about $G_0$. This suggests that the doping atom, which breaks the geometric symmetry of the system, couples the orthogonal wave functions between the electrodes. As a result, the $MR^{FM}$ is reduced by five orders of magnitude and changes its sign from positive to negative as listed in Table 1.

Away from the Fermi energy, the results for a FM 5-ZSiNR are qualitatively similar to those for a FM 4-ZSiNR though quantitative details do occur, see, e. g., the more pronounced DOS peaks for Al-, N-, and P-doped 5-ZSiNRs in the parallel configuration or the increased structure of the spin-down conductance for energies in the range E=0.5-0.6 eV. Interestingly, what distinguishes the FM odd-width 5-ZSiNRs from the FM even-width 4-ZSiNRs is the strongly increased $MR^{FM}$ in doped systems for each X. In the P configuration, the conductance spectra of doped FM 5-ZSiNRs are similar to the corresponding ones of doped FM 4-ZSiNRs as illustrated in panels of the first and the third columns of Fig. 4. In the AP configuration, however, a conductance dip appears at the Fermi energy for doped FM 5-ZSiNRs instead of a conductance peak for doped FM 4-ZSiNRs in all doping cases as shown in the panels of the fourth column of Fig. 4.

The effects of the doping atom on the linear conductance and $MR^{FM}$ of ZSiNRs are summarized in Tables 1 . In Table 1 we compare the conductance and the $MR^{FM}$ of a pristine FM $n$-ZSiNR with that of a B-, Al-, N-, or P-doped FM $n$-ZSiNR in the P and AP configuration for $n=4$ and 5. As can be seen, all $G_P$ are approximately the same, $G_{AP}$ increases by five orders of magnitude while the $MR^{FM}$ reverses direction and decreases by five orders of magnitude in doped 4-ZSiNRs relative to that of the pristine 4-ZSiNR. Similarly, drastic results are obtained for a 5-ZSiNR: here all $G_P$ are approximately the same while the $G_{AP}$ of the pristine ZSiNR is reduced by one to two orders of magnitude when it is edge doped. In contrast, the $MR^{FM}$ does not reverse direction and increases by more than two orders of magnitude. All these significant changes can be used for controlling spin-polarized transport.

Table 1. Linear conductance through pristine and doped FM 4- and 5-ZSiNRs in P and AP electrode configuration and the corresponding magnetoresistance ($MR^{FM}$)

| Dopant | FM 4-ZSiNRs | | | FM 5-ZSiNRs | | |
|---|---|---|---|---|---|---|
| | $G^P$ (μS) | $G^{AP}$ (μS) | $MR^{FM}$ (%) | $G^P$ (μS) | $G^{AP}$ (μS) | $MR^{FM}$ (%) |
| Pristine | 200.0 | 0.02 | $10^6$ | 200.0 | 187.4 | 7 |
| B | 166.9 | 196.0 | −17 | 166.8 | 15.8 | 955 |
| Al | 174.3 | 198.7 | −14 | 180.9 | 7.6 | 2283 |
| N | 182.8 | 197.9 | −8 | 186.9 | 8.2 | 2186 |
| P | 178.8 | 198.5 | −11 | 185.0 | 6.1 | 2948 |

B. *AFM configuration*

In the AFM state, due to the staggered spin polarization of the up and down spins, the sublattice potentials for the up and down spin are also staggered [15,18]. This lowers (increases) the energy of electrons in the π (π*) band and creates an energy gap of approximately 0.2 eV in an AFM 4-ZSiNR or 5-ZSiNR as illustrated in Fig. 4.

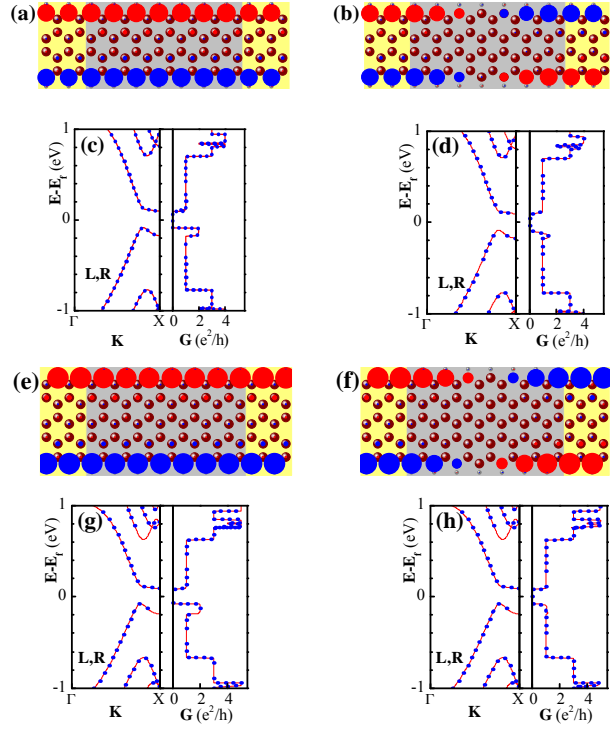

FIG.4. Spin polarization distribution in a two-probe system of a pristine (a) AFM-P 4-ZSiNR, (b) AFM-AP 4-ZSiNR, (e) AFM-P 5-ZSiNR, and (f) AFM-AP 5-ZSiNR. The direction and amplitude of the spin polarization on each atom are indicated by the color (up in red, down in blue) and the radius of the filled circle on the atom sphere, respectively. The corresponding energy bands for the left and right electrodes and the conductance are shown in (c), (d), (g), and (h), respectively. The red solid (blue dotted) curves are for spins up (down).

The atomic distributions of the magnetization of AFM 4- and 5-ZSiNRs in the P and AP configuration are plotted in Fig. 4(a), (b), (e), and (f), respectively. Similar to the FM cases, the magnetization is mainly localized on the edge atoms. In Fig. 4(c), (d) we show the band structure for the left and right electrodes and the corresponding $G$ for an AFM 4-ZSiNR and in Fig. 4(g), (h) for an AFM 5-ZSiNR; (c) and (g) are for the P and (d) and (h) for the AP configuration. The red solid (blue dotted) curves are for spins up (down). As can be seen, neither the band structure nor the conductance is spin resolved. A conductance gap in the range E∈[-0.09 eV, 0.09 eV] appears in the AFM ZSiNRs showing semiconductor characteristics. The conduction band bottom is located at the Brillouin zone edge while the valence band top is not. This results in the conductance peak of height $2G_0$ on the lower edge of the conductance gap in the P configuration as shown in Fig. 4(c) and (g). In the AP configuration the conductance peak is greatly reduced in the 4-ZSiNR, as shown in Fig. 4(d), because the edge states of the same spin in the left and right electrodes are localized on opposite edges of the ribbon. In Fig. 4(h) the conductance peak disappears in the 5-ZSiNR as the ribbon becomes wider. Apart from the suppression of the peak below the gap, the conductance in the AP configuration is quite similar to that in the P configuration.

      The spin degeneracy is broken when ribbons are edge-doped as Fig. 5, for an AFM 4- and 5-ZSiNR makes clear. We also show the corresponding DOSs in the manner of Fig. 3. Again, similar to the FM results shown in Fig. 3, we see the same overall qualitative behavior in Fig. 5 but some quantitative differences as well. Notice that the main gaps in the conductance and the DOS are not affected by the nature of the dopant since the electrodes are made of pristine ZSiNRs and that such gaps are absent in the FM results of Fig. 3. As regards the gap, the conductances of AFM even- or odd- width ZSiNRs do not show much difference compared with those of FM ZSiNRs. In addition, the conductances in the P and AP configurations are almost the same. The main change is the suppression of the conductance peak for spin-down electrons from the P to the AP configuration as shown in Fig. 5. Because the ZSiNRs in the absence (presence) of magnetic field are in the AFM (FM) state and behavior as metals (semiconductors) in the P configuration of electrodes, the magnetoresistance $MR^B$ as defined by Eq. (3) becomes extremely large.

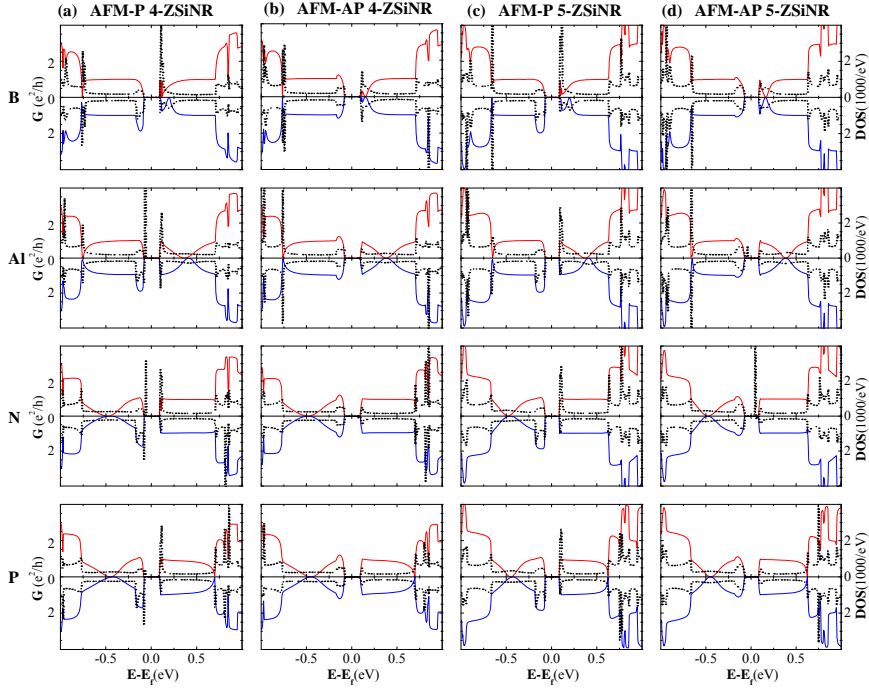

Fig. 5. The same as in Fig. 3 for a 4- and 5-ZSiNR in the AFM state.

## 4. Concluding remarks

We have studied the effect of one substitutionally edge-doping atom in the middle of the scattering region on electron transport through ZSiNRs using the density-functional theory combined with the nonequilibrium Green's-function method. For ZSiNRs in the ferromagnetic state and contacted with antiparallel magnetic electrodes, the doping atom breaks the geometry symmetry and increases the linear conductance of a 4-ZSiNR by five orders of magnitude. In contrast, the formation of bound states around the doping atom strongly decreases the linear conductance of a 5-ZSiNR. This effect results in the suppression of the giant magnetoresistance in 4-ZSiNRs and the appearance of a large magnetoresistance in 5-ZSiNRs, thus showing a strong effect the even or odd width has on it. For ZSiNRs in the antiferromagnetic state, the doping effect is quite limited apart from the removal of the spin degeneracy near the Fermi energy. Conductance dips are introduced by the doping atom due to the formation of localized impurity states. By applying a magnetic field to drive ZSiNRs from the AFM semiconductor to the FM metal state, we can drastically increase the linear conductance.

**Acknowledgments**

This work was supported by the National Natural Science Foundation in China (Grant Nos. 11074182, 91121021, and 11247028) and by the Canadian NSERC Grant No. OGP0121756.